# Developers' Experience with Generative AI - First Insights from an Empirical Mixed-Methods Field Study

Charlotte Brandebusemeyer
Digital Health – Connected Healthcare
Hasso Plattner Institute,
University of Potsdam
Potsdam, Germany
char.brandebusemeyer@hpi.de

Tobias Schimmer
SAP Labs
SAP
Newport Beach, USA
tobias.schimmer@sap.com

Bert Arnrich
Digital Health – Connected Healthcare
Hasso Plattner Institute,
University of Potsdam
Potsdam, Germany
bert.arnrich@hpi.de

**ABSTRACT**

With the rise of AI-powered coding assistants, firms and programmers are exploring how to optimize their interaction with them. Research has so far mainly focused on evaluating output quality and productivity gains, leaving aside the developers' experience during the interaction. In this study, we take a multimodal, developer-centered approach to gain insights into how professional developers experience the interaction with Generative AI (GenAI) in their natural work environment in a firm. The aim of this paper is (1) to demonstrate a feasible mixed-method study design with controlled and uncontrolled study periods within a firm setting, (2) to give first insights from complementary behavioral and subjective experience data on developers' interaction with GitHub Copilot and (3) to compare the impact of interaction types (no Copilot use, in-code suggestions, chat prompts or both in-code suggestions and chat prompts) on efficiency, accuracy and perceived workload whilst working on different task categories. Results of the controlled sessions in this study indicate that moderate use of either in-code suggestions or chat prompts improves efficiency (task duration) and reduces perceived workload compared to not using Copilot, while excessive or combined use lessens these benefits. Accuracy (task completion) profits from chat interaction. In general, subjective perception of workload aligns with objective behavioral data in this study. During the uncontrolled period of the study, both higher cognitive load and productivity were perceived when interacting with AI during everyday working tasks. This study motivates the use of comparable study designs, in e.g. workshop or hackathon settings, to evaluate GenAI tools holistically and realistically with a focus on the developers' experience.

## 1 Introduction

"Generative AI increases productivity" – a phrase frequently encountered in industry and academic contexts. Since the release of AI-powered coding assistants such as GitHub Copilot in 2021, there has been a strong desire to integrate Generative AI (GenAI) into workflows and products. The adoption is driven and encouraged by the promise of increased productivity on the organization, team and individual level. However, Gartner's 2025 Hype Cycle for Artificial Intelligence [23] found that GenAI has entered the "trough of disillusionment" – a period characterized by the recognition that technology does not meet the initial high expectations. Research on developers' interaction with GenAI can contribute to a systematic evaluation of potentials and shortcomings of this new technology.

AI-powered coding assistants like GitHub Copilot were developed to automate programming-related tasks. The aim is to support and relieve programmers and increase their productivity. The focus of research has so far been on evaluating the capabilities and limitations of GenAI regarding output quality [32, 34, 46, 47] and productivity gains [1, 8, 37, 49]. Although technology was developed for programmers, the human aspect and programmers' behavior and experience during the interaction with GenAI are often not the focus of current research. A few recent studies have been approaching this topic [1, 2, 8, 33, 37, 40, 42, 45, 49]. However, empirical studies conducted in real-life organizational settings are limited in academic research [1, 8, 33, 37, 42]. Studies also often rely solely on subjective data from, e.g., questionnaires, interviews, or blog posts [8, 40, 42]. By combining multiple subjective and objective data sources, one can compensate for individual shortcomings of data collection methods and gain a more nuanced understanding of the developer-GenAI interaction. Additionally, there is systematic research missing which examines how different interaction types (in-code suggestions vs. chat prompts), modes (ask, edit and agent), and task types (code generation, debugging, documentation, testing) influence output quality, developers' interaction quality and developers' perceived workload in a natural work environment. Study designs including physiological data gathered via wristbands to analyze developers' GenAI interaction experience more closely are also missing so far.

To address some of these research gaps, we conducted an empirical mixed-methods study with professional software developers at SAP. Combining multimodal data – including subjective experiences via questionnaires, behavioral data from screen, mouse, and keyboard recordings, and physiological data from a wristband – provides a multifaceted impression of the developer-GenAI interaction. The combination of controlled sessions and uncontrolled periods during the study enables both experimental control in a real-world firm setting and a realistic impression of professional software developers' workdays. With this study setup, we investigate how GenAI interactions during (simulated) software engineering tasks influence efficiency, accuracy, and perceived workload.

In this paper, we present the entire study design to demonstrate the feasibility and usefulness of conducting a multimodal study in a firm context with professional software developers. We hope to motivate future studies to employ a comparable study design for an holistic evaluation of developers' interaction and experience with new AI tools. Whilst the entire study setup is described in detail, a focus on certain aspects of the extensive study design and multimodal data gathered needed to be set in this paper: We would like to give first insights into how behavioral and subjective data combined contribute to a deeper understanding of GenAI's impact on developers' experiences and productivity in the software engineering context. Analyses focus on the controlled sessions of the study, whilst also providing first insights into results from the uncontrolled study period.

The following research questions are going to be addressed:
How does GenAI interaction impact the productivity indicators…

    **(1)** … efficiency

    **(2)** … accuracy

    **(3)** … the developer experience indicator perceived workload

during simulated software engineering tasks?

This study contributes to emerging research that takes a developer-centered view on the developer-GenAI interaction in a real-world firm setting.

## 2  Related Work

Generative AI (GenAI) refers to AI technology that learns from data to autonomously generate new, meaningful, and contextually appropriate content across various applications [10]. GitHub Copilot - from here on referred to only as Copilot - is one of several GenAIs that can assist developers by generating, completing and modifying programming code based on the context of the codebase and natural language prompts. It is integrated in software development environments (IDEs) like VS Code and can function as an "AI pair programmer" for developers.

In recent years, multiple studies have been conducted to evaluate Copilot's output quality and developers' productivity gains through AI usage. Based on benchmark tasks, Dakhel et al. [32] found that Copilot can generate solutions for nearly all given tasks, but its outputs are less often correct than those of humans, making it a potential asset for experienced developers but a liability for novices who may not detect non-optimal suggestions. So far, there is no consensus on how to best measure software developers' productivity, but time savings, acceptance rate of Copilot suggestions, and successful completion of predefined tasks are frequently considered metrics to evaluate developer productivity in the context of AI-assisted programming. Empirical studies using these metrics report mixed results: Bakal et al. [1] found that developers using Copilot accepted 33% of suggestions and had a 20% increase in lines of code; Peng et al. [37] observed that programmers completed an HTTP server task 55.8% faster with Copilot, with the largest productivity gains among less experienced programmers; while Vaithilingam et al. [45] found no significant improvement in task time or success, though participants still valued Copilot as a useful starting point for daily programming tasks. In this study, we consider task duration, the number of Copilot's suggestions, and successful task completion as productivity metrics. Additionally, we examine the impact of Copilot interaction types, developers' interaction intensity, and software development task categories on these metrics. This data is also combined with subjective workload ratings from the developers to gain a more nuanced analysis of not solely the output quality and productivity gains of Copilot, but also the interaction behavior and experience with Copilot.

Several studies in the software engineering context have demonstrated benefits of using mixed-methods approaches that combine objective telemetry, physiological measures and subjective data to provide deeper insights into developers' experiences during work [6, 17, 36]. However, only a few studies have used mixed-methods study designs and multimodal data to analyze the developer-GenAI interaction. For example, Tang et al. [44] combined IDE telemetry, eye-tracking, and subjective workload measures to study how developers validate and repair LLM-generated code. They showed that knowing whether the code was AI-generated influenced both the developers' debugging behavior and cognitive load, with cognitive load being higher when developers were aware that the code was AI-generated. Ziegler et al. [49] combined survey data, based on the SPACE framework [13], with telemetry data to assess the impact of Copilot on professional developers' productivity, finding that the acceptance rate of in-code suggestions correlates with perceived productivity. With the study presented in this paper, we contribute to the growing mixed-methods and multimodal research on developer-GenAI interaction. The aim is to gain a more developer-centered perspective on developers' perception and productivity during their work with AI.

## 3 Study Procedure

Each participant completed a four-day study, with controlled sessions on the first and last days and an uncontrolled period in between (Figure 1). Physiological data were continuously recorded using the EmbracePlus wristband. Subjective experiences were assessed through questionnaires, and behavioral data were captured via screen, mouse, and keyboard activity during the controlled sessions (Figure 1, Data recorded). The study took place at two SAP locations in the US.

### 3.1 Controlled sessions

Each participant met the experimenter individually in a meeting room in the firm for the controlled sessions that framed the study (first controlled session on the first day, last controlled session on the last day). Here, they were briefed on the study procedure, data collection, and data privacy, and then gave their informed consent to participate. The participant was then equipped with the EmbracePlus wristband to measure the physiological activity, and the screen, mouse and keyboard recording was started on the laptop. The session began with a pre-questionnaire covering job characteristics, programming and GenAI experience, work satisfaction, developer experience, and a personality test (Figure 1, Pre-Questionnaire).

Next, participants completed an n-back task (1-back to 3-back) to assess cognitive performance. Alphabetical letters needed to be memorized n positions back. After each difficulty level, the cognitive load and the stress level were rated. A startle cue (honking sound) was introduced after the 3-back task to control for physiological responses during later data analysis (Figure 1, Cognitive n-back tasks with a startle event). A five-minute breathing meditation video then served to relax the participant and gave time for physiological activity to return to a resting state (Figure 1, 4-7-8 breathing meditation).

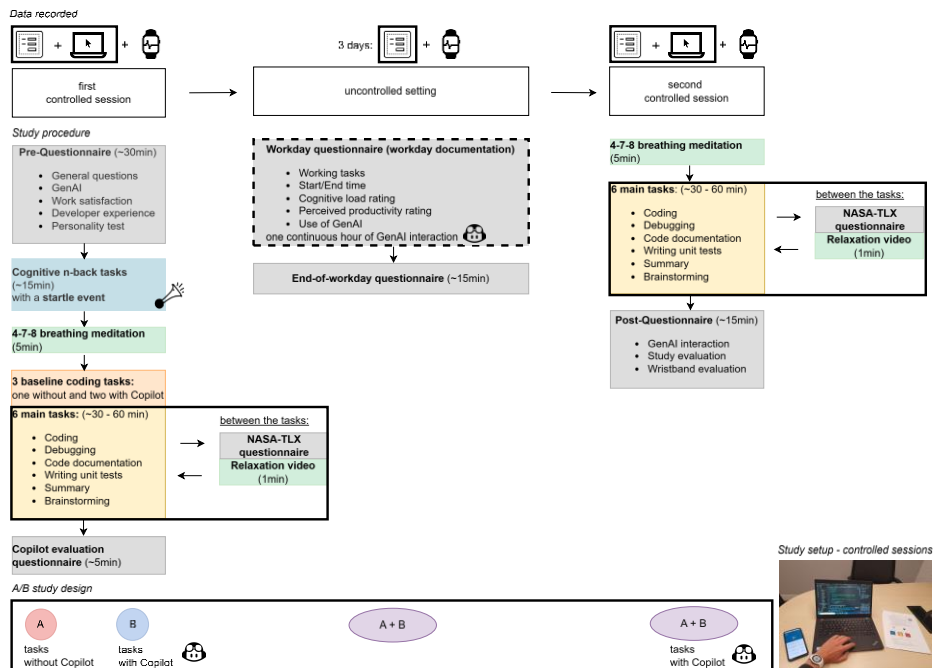

**Figure 1: The study procedure and setup together with the A/B study design is depicted. In this paper, the focus lies on the black surrounded sections.**

Participants then completed three baseline coding tasks in Java—one without Copilot and two with—to familiarize themselves with the Visual Studio Code IDE and the Copilot integration. After each task, participants filled in a NASA Task Load Index questionnaire (NASA-TLX), followed by a one-minute relaxation video. A two-minute video was played for further relaxation before the main tasks (Figure 1, 3 baseline coding tasks).

The main phase included six randomized tasks: coding, debugging, code documentation, writing unit tests, summary and brainstorming tasks. As before, the NASA-TLX questionnaire was filled in, and a one-minute relaxation video was shown between tasks (Figure 1, 6 main tasks). In the A/B design for the first controlled session, half of the participants did not use Copilot during the main tasks; the others did (Figure 1, A/B study design). Participants completed a Copilot evaluation questionnaire either after the baseline tasks (if not using Copilot) or at the end of the session (if using Copilot) (Figure 1, Copilot evaluation questionnaire).

The second session mirrored the first, starting with a meditation video, followed by equivalent tasks, but this time performed by all (both groups) with Copilot (Figure 1, A/B study design). After the tasks, participants filled in a post-questionnaire with final evaluations of their GenAI interaction during the study and an evaluation of the study procedure and setup (Figure 1, Post-Questionnaire). Upon completion of the study, each participant received a 50$ Amazon voucher via email as compensation for their time and effort in participating.

## 3.2 Uncontrolled setting

During the three days between the two controlled sessions, the participants followed their normal workdays whilst documenting their working tasks. For each task, they tracked the start and end times, the task, their feeling of cognitive load and productivity, and use of GenAI (if any). This workday questionnaire was filled in on a sheet of paper. Participants were encouraged to use a GenAI tool of their choice for at least one continuous hour per day to ensure a consistent time window for physiological data analysis (Figure 1, Workday questionnaire). At the end of the workday, the participants filled in an end-of-workday questionnaire assessing their developer experience and GenAI interaction for that day (Figure 1, End-of-workday questionnaire).

# 4 Study Design and Technical Details

## 4.1 Study design

*4.1.1 A/B testing.* The study used an A/B test design: Group A completed the first session without Copilot and the second session with Copilot, while Group B used Copilot in both. This setup allows for both within-group and between-group comparisons of GenAI interaction over time (Figure 1, A/B study design).

*4.1.2 n-back task.* The n-back task, commonly used in Cognitive Psychology and Neuroscience, assesses working memory, cognitive load, and attention. Participants viewed a sequence of letters, each shown for 0.5 seconds, followed by a cross for two seconds. The task was to identify whether the current letter matched the one n steps earlier. Difficulty increased with n, since more letters need to be stored in working memory. The task was performed three times with increasing difficulty levels (n = 1, 2 and 3). 48 letters were shown per task. The number of errors was displayed as feedback to the participants to maintain their motivation. After each difficulty level, participants rated their cognitive load and how insecure, discouraged, irritated, stressed and annoyed they were on a 7-point Likert scale (1 = very low, 7 = very high) (Figure 1, Cognitive n-back tasks with a startle event).

*4.1.3 Startle event.* A startle event in the form of a honking sound was played after the final letter of the 3-back task as a control cue to verify the reliability of the physiological data recording (Figure 1, Cognitive n-back tasks with a startle event).

*4.1.4 Relaxation videos.* Relaxation videos were used between tasks to help participants recover physiologically after a cognitively demanding task before completing the next one (Figure 1, 4-7-8 breathing meditation, Relaxation video). Following the startle event, a five-minute guided meditation breathing video [22] using the 4-7-8 breathing technique promoted relaxation. Between tasks, one-minute nature scene videos were shown to the participants with the note to breathe deeply and let the mind dive into the scene in the video. The aim was to minimize mind wandering whilst at the same time using the positive effect of nature on mental well-being [4] to relax the participants and reduce their physiological activity between the tasks. Video material for the generation of the relaxation videos was copyright-free [14–16, 22, 39]. The open-source video editor Shotcut was used to create the video snippets [29].

*4.1.5 Tasks during the controlled sessions.* The programming-related tasks for the controlled sessions were selected from the HumanEval-X benchmark dataset [48]. The coding language was Java. To make the task difficulty comparable between different tasks and the two controlled sessions, code metrics were used to select 15 tasks of comparable cognitive demand. Specifically, the cognitive complexity [7], cyclomatic complexity [28], Halstead metrics [20], lines of code (LOC), non-comment LOC (NCLOC), comment LOC (CLOC) and nested block depth (NBD) were considered. The metrics were calculated by the MetricsReloaded plugin [25] in IntelliJ IDEA 2024.3.3.

Three baseline tasks in the first session served to familiarize the participants with the IDE and Copilot and were excluded from later analyses (Figure 1, 3 baseline coding tasks). The 12 main tasks spanned six categories - coding, debugging, documentation, unit testing, summaries, and brainstorming - with each category included in both sessions. The summary tasks consisted of summarizing an email and a video-recorded meeting; brainstorming tasks asked for ideas on team-building activities and how to make meetings more engaging, collaborative and efficient (Figure 1, 6 main tasks during first and second controlled session). For every participant, the order of the 12 tasks was randomized, with each task category occurring once during each controlled session. This way, order, learning, fatigue, and other sequence-related biases are controlled for.

*4.1.6 Questionnaires.* The participants' subjective experience was evaluated via six questionnaires. Most questions are based on previous research for comparability reasons.

*4.1.6.1 Pre-Questionnaire.* The pre-questionnaire was filled in at the beginning of the first controlled session (Figure 1, Pre-Questionnaire). Questions regarding the work environment [41], work experience [41], GenAI use and satisfaction [24, 30], work satisfaction [38, 41], developer experience according to the DevEx framework [11] - expanded with further questions on the flow state [31] and the DX Core 4 framework [43] - and the 50-item International Personality Item Pool (IPIP) based on Goldberg's markers for the Big-Five factor structure [18, 19] were filled in. The aim was to gain general yet extensive insights into the participant pool of this study.

*4.1.6.2 NASA Task Load Index.* The NASA-TLX is a widely used questionnaire to assess perceived workload [21]. It consists of six 21-point Likert-scaled questions asking for mental demand, physical demand, temporal demand, performance, effort and frustration during a task. Participants filled in the questionnaire after each controlled session task to capture and later compare task impact (Figure 1, NASA-TLX questionnaire). The raw NASA-TLX score – the average of the six ratings – provided the measure for the perceived workload in this study.

*4.1.6.3 Copilot Evaluation.* After using Copilot in the first controlled session (post-baseline for Group A, end of first session for Group B), participants' interactions with in-code suggestions and chat were evaluated separately (Figure 1, Copilot evaluation questionnaire). The questions are based on the SPACE framework [13] and were adapted to evaluate specifically GitHub Copilot use.

*4.1.6.4 Workday Questionnaire.* During the normal workday, participants used a paper questionnaire to log tasks, start and end time of each task, cognitive load (7-point Likert scale), perceived productivity (6-point Likert scale), and whether GenAI was predominantly used to solve the task. They were encouraged to use some form of GenAI for at least one consecutive hour during their workday (Figure 1, Workday questionnaire). The aim was to evaluate for which tasks developers currently use GenAI during their normal work routines and how useful they find the interaction.

*4.1.6.5 End-of-workday Questionnaire.* Similar questions to the pre-questionnaire were asked at the end of each workday [11, 38, 41]. This questionnaire was tailored to capture participants' experiences and extraordinary occurrences that might affect data analysis during one specific day (Figure 1, End-of-workday questionnaire).

*4.1.6.6 Post-Questionnaire.* The post-questionnaire [30, 41], completed after the second controlled session, evaluated participants' GenAI interaction, wristband use, and the study setup to gather insights for future studies (Figure 1, Post-Questionnaire).

*4.1.7 GenAI use.* GitHub Copilot is integrated in the software development environment (IDE) VS Code. It gives developers a broad range of interaction possibilities: Participants could freely switch models, interaction types (in-code suggestions, chat, in-line chat), and modes (ask, edit, agent) in this study design. Copilot selects an appropriate prompt template based on the chosen interaction type and mode. Differing templates explain differing model outputs. The three main interaction types differ in the location where code suggestions are generated, the context scope of the codebase they consider, response latencies and intended use. In-code suggestions appear at the position of the cursor, using only a few lines of code above and below the cursor position as a context for code completions. This interaction type is optimized for short latencies, which enables real-time suggestions similar to autocompletion, but generates more lines of code with ideally more contextual relevance. In-line chat is used within the codebase for specific questions and adaptations in the code. The considered context is also limited to the surrounding lines of code. For chat, a separate window enables a dialogue-like interaction for an ongoing conversation. Chat considers a wider context of the codebase within the range of an entire file or multiple files. Responses are optimized for creativity and explanations. GPT-4o was set as the default model, and the "ask" mode was the default mode for the chat interaction at the beginning of the controlled sessions. During the uncontrolled period, they were free to use any GenAI tool. The study aims to examine the developers' interaction with GenAI tools, rather than assessing specific tools or model performances.

## 4.2 Technical Setup

*4.2.1 Screen, mouse and keyboard recording.* Behavioral data during the controlled session were gathered via screen, keyboard and mouse recordings. The open-source Open Broadcast Software (OBS) was used for screen recordings [35]. Keyboard use and mouse movement were recorded via the JNativeHook library for global keyboard and mouse listening for Java [3]. Recordings were limited to the study laptop and the controlled sessions for personal and firm data privacy reasons.

*4.2.2 Wearable device.* Physiological data was recorded via the EmbracePlus wristband (FDA-cleared and CE-certified [9]). Heart activity (PPG), electrodermal activity (EDA), skin temperature, and wrist movement were recorded during the whole study period. The device provides raw, unprocessed physiological data for transparent and explainable data analysis.

## 4.3 Participants

Twenty-two SAP employees participated in the study - fourteen from one SAP location, eight from another, both within California, US. The participants consisted of 12 software developers/engineers, 8 senior software developers/engineers, one senior quality specialist and one principal software architect, averaging 5-10 years of IT/programming experience. All had Java experience (a study requirement), with Java (n=22), Python (n=12), and JavaScript (n=11) as the most used programming languages among participants. Based on a 5-point scale, participants were on average highly proficient in Java (M=4.45, min=3, max=5) and also in GenAI use (M=3.77, min=2, max=5). The participants had been using GenAI at work for on average 6-12 months and for at least 1-6 months. The most commonly used GenAI tools at work were GitHub Copilot and ChatGPT. In their free time, the participants mostly use ChatGPT and Google Gemini, with two participants stating that they do not use GenAI in their free time. For one participant, only the pre-, post- and copilot evaluation questionnaire data could be used, since instructions were not followed. We refrained from collecting age and gender information due to privacy concerns. Participation in this study was voluntary, and the design and procedure of the study were reviewed and approved by the ethics committee of the University of Potsdam.

## 4.4 Data Preparation and Analysis

In this paper, the focus lies on analyzing the two controlled sessions of the study where developers work on simulated software engineering tasks. Subjective experiences (from the NASA-TLX questionnaires) and behavioral interactions with GenAI (from the screen recordings) contribute to the evaluation. Initial insights into participants' use of GenAI during the uncontrolled part of the study (during their workday) are also given.

*4.4.1 Data labeling.* To analyze participants' interaction with GenAI during the controlled sessions, ~66 hours of screen recordings were manually labeled. Labels included per participant and per task were: controlled session day (first or second day), task order (1-6 per day), task duration, task completion (completed, partly completed/not completed), Copilot use (with or without), the number of in-code suggestions and the number of chat prompts. The in-line chat editor was categorized as chat interaction in this study. The labeling of edge cases was reviewed by all authors of this paper.

*4.4.2 Statistical Methods.* Significant within-participant differences were assessed using the Wilcoxon signed-rank test, and between-participant comparisons were conducted with the Mann–Whitney U test.
To predict task completion, duration, and perceived workload, (generalized) linear mixed-effects models were used, with task categories and Copilot interaction types as fixed effects. For task duration, the number of in-code and chat interactions were additionally added as fixed effects to reflect interaction intensity. Participants and tasks were included as random effects to account for individual and task-level variability. Details on constructed mixed-effects models, model fit evaluations, and the associated post hoc tests and p-value adjustments can be found in the supplemental material [5].

*4.4.3 Pre-analyses for evaluating the controlled sessions.* Pre-analyses found no statistically significant difference in task durations between the two controlled sessions (first and last day of the study) nor between the two participant groups (group A not using GenAI during the first controlled session vs. group B using GenAI during both controlled sessions) (Figure 1). Therefore, the tasks from both controlled sessions and the two participant groups are analyzed together in the following. Please refer to the supplemental material for details on the performed analyses [5].

## 5 Results

The following analysis examines how GenAI/Copilot use impacts the productivity indicators task duration (efficiency), task completion (accuracy), and raw NASA-TLX scores (perceived workload) during the predefined tasks of the controlled sessions.

## 5.1 Task duration (efficiency)

To assess the impact of Copilot use on task duration, linear mixed-effects models predicting the log-transformed task duration were fit on the data and compared:
log(task duration) ~ interaction type + num_inCode + num_chat + task categories + scaled raw NASA-TLX + (1|participants) + (1|tasks)

The impact of interaction types (without GenAI use, only in-code suggestions, only chat prompts or both in-code suggestions and chat prompts), interaction intensity (number of in-code suggestions (num_inCode) and number of chat prompts (num_chat), task categories (coding, debugging, documentation, unit testing, summary brainstorming) and perceived workload (scaled raw NASA-TLX scores) on task duration is analyzed in the following paragraphs. As detailed in the supplemental material [5], both interaction type and intensity significantly improve model fit, individually and combined. This suggests that how and how much developers used Copilot had an impact on the task durations.

*5.1.1 Impact of interaction type on task duration.* To examine the effect of interaction type on task duration, pairwise comparisons using a Tukey post hoc test were conducted. Participants working on the tasks with only chat (estimate=0.44, SE=0.120, p=0.0016) or only in-code suggestions (estimate=0.46, SE=0.159, p=0.0221) were significantly faster than without Copilot. When both in-code suggestions and chat were used during a task, task duration did not differ significantly compared to not using Copilot (estimate=0.223, SE=0.170, p=0.5559). No significant differences could be observed between the three Copilot-assisted conditions (Figure 2). The results suggest that while individual interaction with either in-code suggestions or chat improves efficiency, combining interaction types does not yield additional time savings compared to not using Copilot to solve a task.

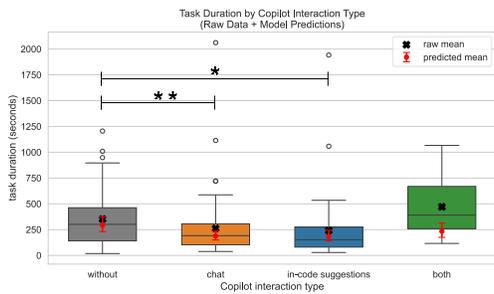

**Figure 2: Raw task durations and model-predicted durations grouped by Copilot interaction type are shown. Asterisks denote statistical significance (* p≤0.05; ** p≤0.01).**

*5.1.2 Impact of interaction intensity on task duration.* To assess how interaction frequency/intensity affects efficiency, average task durations were compared across interaction types and counts against a no-Copilot baseline at 342.6s (Figure 3). Both in-code suggestions and chat prompts reduced task duration at lower interaction counts, with in-code showing consistent efficiency gains up to 13 interactions, and chat suggestions up to six interactions. Beyond these levels, task durations mostly exceeded the no-Copilot baseline. Combined use of both interaction types showed slight efficiency gains at low interactions but decreased performance at higher levels. Longer task durations at a higher number of chat prompts may reflect time spent refining prompts. In contrast, the relative efficiency of in-code suggestions may result from faster execution on a trial-and-error basis. Combined use of interaction types mostly indicated unsatisfying results from one mode, leading to a switch. In sum, moderate GenAI interactions improve efficiency, while excessive interactions may introduce overhead that diminishes benefits.

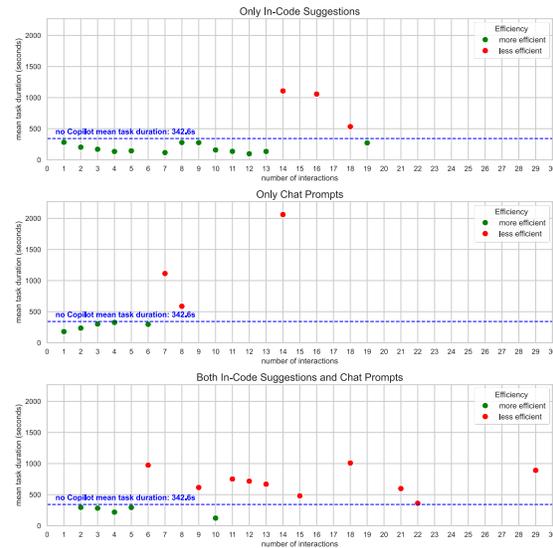

**Figure 3: These three diagrams compare the efficiency of using Copilot compared to no Copilot (blue dashed line) during tasks. The three plots represent the three Copilot interaction types. Dots show the mean task duration per number of interactions (green dots: more efficient than not using Copilot; red dots: less efficient).**

*5.1.3 Impact of task categories on task duration.* Next, the influence of the task categories on task duration was analyzed. A full linear mixed-effects model including task categories as a predictor fit significantly better to the data than a reduced model excluding the predictor ($\chi^2(5)=20.04$, $AIC_{full\_model}=388.29$, $AIC_{null\_model}=398.32$, p=0.0012). This indicates that task duration varied by task type.

In the full model with the coding tasks as the reference, debugging tasks were associated with significantly longer durations (estimate=0.6606, p=0.0152), as were summary tasks (estimate=0.4887, p=0.0494). No significant differences were found for the other tasks. Post-hoc pairwise comparisons (Tukey-adjusted) indicated marginal differences between debugging vs. documentation (p=0.0526) and vs. brainstorming tasks (p=0.0592), with debugging tasks taking longer. No other task pairs differed significantly. The results suggest that debugging tasks, and to a lesser extent summary tasks, required more time than other task categories.

Perceived workload, measured via the scaled raw NASA-TLX score was a significant positive predictor of task duration

(estimate=0.32, p<0.001). An increase in perceived workload was associated with increased task duration.

In sum, both the interaction type and intensity of Copilot use significantly influenced task duration. Moderate use of either chat or in-code suggestions improved efficiency compared to no Copilot use, while excessive and/or combined interactions diminished the effect. Task type and perceived workload also help predict duration, with debugging tasks taking longer, and a higher workload associated with increased task duration.

### 5.2 Task completion (accuracy)

A generalized linear mixed-effects model was used to assess how GenAI interaction types and task categories influence task completion:

task completion ~ interaction type + task categories + (1|participants) + (1|tasks)

*5.2.1 Impact of interaction type on task completion.* Interaction type was a marginal predictor for task completion ($\chi^2(3)$=7.4417, $AIC_{full\_model}$=223.91, $AIC_{null\_model}$=225.35, p=$p_{adjusted}$=0.0591), with only chat prompts significantly increasing task completion likelihood compared to no GenAI (estimate=1.19, p=0.0078). In-code suggestions or combining both interaction types showed no significant effect. An exploratory Tukey post-hoc test confirmed that using only chat prompts led to significantly higher task completion than no Copilot use (estimate=1.19, SE=0.448, p=0.0389). Thus, using chat alone appears to be most beneficial compared to not using Copilot. The brainstorming and summary tasks could only be completed with either no Copilot interaction or chat interaction, but not via in-code suggestions, which should be considered when interpreting the results.

*5.2.2 Impact of task category on task completion.* Task categories significantly influenced task completion prediction ($\chi^2(5)$=21.69, $AIC_{full\_model}$=223.91, $AIC_{null\_model}$=235.6, p<0.001, $p_{adjusted}$=0.0012). Compared to coding, debugging (estimate=-2.92, SE=0.9539, p=0.0022), testing (estimate=-3.25, SE=0.9503, p<0.001), documentation (estimate=-2.32, SE=0.9506, p=0.0144) and summary (estimate=-2.96, SE=0.9713, p=0.0023) tasks had significantly lower task completion rates. The Tukey Post-hoc test confirmed that task completion rates for debugging (p=0.0293), testing (p=0.009) and summary (p=0.0300) were significantly lower than for coding, while the difference for documentation was not significant after Tukey adjustment (p=0.1403). Since all brainstorming tasks were completed successfully, there was a complete separation in the data, leading to model instabilities. Nevertheless, we could observe that the task category significantly influenced task completion rates, with coding and brainstorming tasks showing the highest task completion rates (Figure 4).

In sum, task completion was influenced by both the type of Copilot usage and the task category, with using only chat-based prompts and working on coding or brainstorming tasks each being associated with higher success rates in this study.

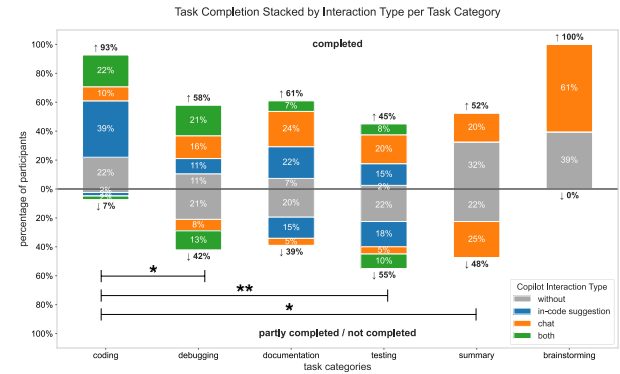

**Figure 4: The percentage of tasks completed (or not) via a specific Copilot interaction type is depicted (stacked bars). The percentage of task categories completed vs. partly completed/not completed by the participants is presented above and below the bars. Asterisks denote statistical significance (\* p≤0.05; \*\* p≤0.01).**

### 5.3 Perceived workload

A linear mixed-effects model was fit to predict participants' perceived workload whilst working on the tasks:

log(raw NASA-TLX + 1) ~ interaction type + task categories + (1|participants) + (1|tasks)

*5.3.1 Impact of interaction type on perceived workload.* Interaction types significantly impact the prediction of perceived workload ($\chi^2(3)$=32.02, $AIC_{full\_model}$=403.01, $AIC_{null\_model}$=429.03, p=$p_{adjusted}$<0.001). Compared to not using Copilot, using only chat prompts (estimate=−0.41, SE=0.1045, p<0.001) and only in-code suggestions (estimate=−0.60, SE=0.1220, p<0.001) were both associated with significantly lower reported workload. The combined use showed no significant difference from not using Copilot (estimate=−0.13, SE=0.1376, p=0.3338). Post-hoc pairwise comparisons using Tukey adjustment confirmed that both chat prompts (p<0.001) and in-code suggestions (p<0.001) conditions resulted in lower perceived workload ratings than without using GenAI. Additionally, only using in-code suggestions resulted in significantly lower perceived workload compared to the combined GenAI conditions (p=0.0064) (Figure 5). No other pairwise comparisons were significant.

In sum, both in-code suggestions and chat prompts lowered perceived workload compared to not using GenAI. However, using both together led to a higher workload than using in-code suggestions alone, suggesting that switching between interaction types may increase workload compared to focusing on a single interaction type.

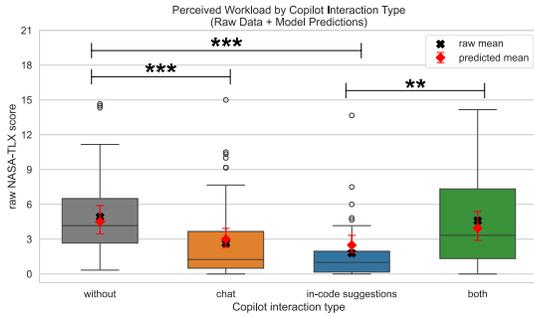

**Figure 5: The raw NASA-TLX scores and model predictions of perceived workload grouped by Copilot interaction type are depicted. Asterisks denote statistical significance (** p≤0.01; *** p≤0.001).**

*5.3.2 Impact of task category on perceived workload.* Task categories significantly improved the prediction of perceived workload ($\chi^2(5)$=11.3, $AIC_{full\_model}$=403.01, $AIC_{null\_model}$=404.31, p=$p_{adjusted}$=0.0458), though no individual categories differed significantly from coding. Comparability between the cognitive load of the tasks was intended in the study design by choosing tasks from the HumanEval-X dataset and selecting tasks based on cognitive load code metrics. These results validate the selection process of the tasks.

Overall, using either in-code suggestions or chat prompts lowered perceived workload, whilst combining both stayed comparable to not using GenAI or increased it compared to in-code suggestions. Task categories influenced the overall perceived workload, but had no significant individual effects.

> Moderate use of either in-code or chat suggestions improves efficiency and reduces workload, while excessive or combined use lessens these benefits. Only using chat boosts task completion in this study. Task type affects efficiency and task completion, with coding and brainstorming tasks having higher success in this study.

*5.3.3 Perceived cognitive load and productivity during the uncontrolled setting.* Out of 445 documented working tasks, 120 (27.42%) involved AI interaction. Of the 22 participants, 12 found AI helpful every time they used it, one never found it helpful, one did not use AI in the uncontrolled setting, and the remaining eight participants found AI helpful only sometimes. Comparing tasks worked on with vs. without AI interaction, cognitive load ratings differed significantly, with higher ratings for tasks with AI interaction compared to without (W=0, p<0.0001, r=-0.8760). Productivity ratings also differed significantly and were higher when interacting with AI compared to without (W=26, p=0.002, r=-0.6788). Within the subgroup of eight participants who had mixed evaluations of AI helpfulness during their tasks, neither cognitive load nor productivity ratings differed significantly between tasks rated as helpful vs. not helpful (cognitive load: W=7, p=0.2367, r=-0.8181; productivity: W=10, p=0.499, r=-0.7930).

## 6 Discussion

**Copilot interaction vs. no interaction.** This study found that using one interaction type, i.e. either in-code suggestions or chat, can increase efficiency (shorter task duration) and reduce perceived workload (NASA-TLX score) compared to not using Copilot. These findings align with Peng et al.'s [37], who also found that task completion was faster with Copilot than without it. A notable result of this study is that combining both interaction types does not yield benefits in task duration or perceived workload. Switching between interaction types may indicate that the initially chosen interaction type was suboptimal to solve a given problem. Such switches may resemble small interruptions that involve a shift in the problem-solving approach and of attention, which can be associated with increased workload [26, 27].

During the study days, when participants documented their tasks and AI use in an uncontrolled work setting, they reported a higher cognitive load during tasks involving AI interaction than during non-AI tasks; yet, they also felt more productive when using AI. Among participants with mixed views on AI helpfulness, experienced cognitive load and productivity did not differ between tasks where AI was perceived as helpful vs not helpful. Comments from the workday documentation and the results from the controlled sessions suggest that the elevated cognitive load during AI-related tasks stems from switching between coding-related tasks (e.g., while working on code in an everyday setting, such as coding, debugging, and testing which alternate naturally) and different interaction types with AI. Tang et al. [44] similarly found in their study that when developers knew that code was generated with AI, cognitive load was perceived higher. They observed frequent switching between source code and comments during validation and repair of LLM-generated code. Their results strengthen our interpretation that higher cognitive load during AI-related tasks may be higher due to increased context switching. Further detailed analyses on the relation between working tasks and AI use are going to be addressed in future studies.

**In-code suggestions vs. chat.** The degree to which GenAI use has beneficial efficiency effects depends on the interaction intensity, i.e. the amount of GenAI use. High use of in-code suggestions can still lead to efficiency gains, whereas only a low number of interactions with chat or a combination of both interaction types proves more efficient than not using Copilot. As could be observed in the screen recordings, in-code suggestions are used in a trial-and-error fashion, leading to a higher volume of generated and discarded code. In contrast, the interaction with chat requires more high-level thought to formulate the prompt, explaining the lower number of chat interactions for efficiency gains. If a prompt is formulated in a comprehensive way for the model to generate the desired output, fewer interactions are needed to receive a result. Both interaction types individually increase efficiency up to a certain number of interactions compared to not using Copilot whilst the combination of both interaction types during a task lessens the benefits. The study results also indicate that perceived workload is lowest for in-code suggestions, which can be explained by the code suggestions appearing directly at the cursor position. The benefit of chat for task completion (accuracy) may be explained by the interaction type considering more code context when giving an output.

**Task categories impact perceived workload in general and differ in their impact on efficiency and accuracy.** For the controlled sessions, code-related tasks were selected from the HumanEval-X dataset based on code metrics to keep the tasks across task categories comparable. Task categories did not differ in perceived workload, which validates the selection process of the tasks based on code metrics associated with cognitive load. Nevertheless, they have an impact on perceived workload. Results further indicated that more time was required to complete debugging tasks, and coding and brainstorming tasks had higher successful completion rates. Task categories, therefore, in general impact perceived workload and differ in their impact on efficiency and accuracy.

## 7 Considerations, Limitations and Outlook

The question may arise how transferable these study results and their interpretation are for GenAI interaction in general with other tools. Whilst interaction behavior and experience are tool-specific, GenAI coding assistants either have or are moving towards including both in-code suggestions and a natural language or chat interface. This study has shown that the way of interacting with coding assistants, as in the interaction types used, can impact developers' experience and productivity. We therefore expect transferability of our results, but more studies in this area are needed for verification.

The aim of the study was to capture developers' interaction with GenAI as realistically as possible. While a natural work environment increases realism, one cannot control for all confounding factors. By including both controlled and uncontrolled periods, along with questions regarding mood, work environment and unusual occurrences, we accounted for some of these factors to contextualize the results. Adding to the realistic setting, participants were proficient and professional Java developers with varying levels of GenAI experience. Future studies could verify these study results with a different developer cohort regarding programming language, programming experience, and GenAI experience.

The controlled sessions involved simpler tasks than developers' daily software engineering tasks. Simplifications were made to manage time constraints of the study, ensure consistent task length and workload across tasks, whilst also considering a variety of task types. During the tasks, the participants used a laptop with a German (QWERTZ) keyboard, unfamiliar to the US-based developers. Participants were, however, informed about the layout differences and were given an introduction to the altered key placements. All tasks and each interaction type included typing; therefore, any additional time, cognitive effort or frustration would affect all conditions similarly. No task or interaction type bias is expected. Task randomization and aggregated analyses across participants further minimize keyboard-related effects.

The NASA-TLX performance subscale is inversely scaled compared to the others, which some participants only noticed after completing several tasks. They reported the issue and the scores were manually adapted after the study. To maintain consistency with the standard NASA-TLX format, the questionnaire structure was not modified in advance.

Future work is going to analyze developer–GenAI interaction during the uncontrolled period in greater detail and examine developers' cognitive load using physiological, keyboard, and mouse data. A similar study setup could be used to study developers' interaction with different and/or multiple code generation tools and large language models to understand the opportunities and risks of GenAI for cognitive load in software development. Risks include additional interruptions and generated code that is difficult to review and maintain [12], impacting developer-AI interaction.

## 8 Conclusion

In this study, we examined professional developers' experience during their interaction with GenAI in their everyday work setting. The focus of this paper was to analyze the controlled sessions of the study and consider the efficiency, accuracy and perceived workload, as well as the impact of different interaction types with Copilot. We found that both interaction types used individually positively affect efficiency and perceived workload compared to not using Copilot. Chat use also improved task accuracy, compared to no Copilot use. However, combining both interaction types did not show additional benefits, which is likely due to the increased workload resulting from interaction and context switching. The findings highlight the need for a more nuanced understanding of how developers interact with GenAI. Mixed-methods study setups like the one described in this paper enable a holistic analysis of developers' interaction experience with GenAI tools within an everyday work environment. In future, similar study setups that combine subjective and objective data sources could be deployed within firms for detailed evaluation of GenAI tools and the experienced interaction of developers with them.

## SUPPLEMENTAL MATERIAL

Details on methods and analyses are made available [5].

## ACKNOWLEDGEMENTS

The work of Charlotte Brandebusemeyer is funded by the SAP-HPI Research Program.